\documentclass[epj]{svjour}

\usepackage{graphicx}
\usepackage{fancyhdr}
\def\makeheadbox{{
 }}
\def\mytitle{My title} 
\def\myauthors{My name}  
\def\mytype{My type of session}
\def\mysession{My session}

\def\mytitle{Anticipating a New Golden Age} 
\def\myauthors{Frank Wilczek}    
\def\mytype{Review}
\def\mysession{\myauthors}

\newcommand{\doefunds}{This work was supported in part by 
funds provided by the U.S. Department of Energy under cooperative research agreement
\#DF-FC02-94ER40818.}
\begin{document}
\title{Anticipating a New Golden Age\footnote{Based on an invited talk at SUSY07, Karlsruhe, July 2007.  This paper follows the talk closely.}}
\author{Frank Wilczek
\thanks{\emph{Email:} wilczek@mit.edu}%
}                     
%
%
\institute{Center for Theoretical Physics, Department of Physics, Massachusetts Institute of Technology, Cambridge, MA 02139
}
%
\date{}

\abstract{
The standard model of fundamental interactions is remarkably successful, but it leaves an unfinished agenda.   Several major questions seem ripe for exploration in the near future.   I anticipate that the coming decade will be a Golden Age of discovery in fundamental physics.
} 
\maketitle

\section{Where We Stand}

\subsection{Celebrating the Standard Model}

At present, the standard model of particle physics stands triumphant.   It has survived testing far beyond the range of energies for which it was crafted, and to far greater precision.   

Even the ``ugly'' parts look good.
Unlike the gauge part of the standard model, whose parameters are few in number (namely, three) and have a beautiful geometric interpretation, the part dealing with fermion masses and mixings contains many parameters (about two dozen in the minimal model) that appear merely as abstract numbers describing the magnitudes of Yu\-kawa-type couplings of one or more hypothetical Higgs fields.    In the present state of theory all these numbers must be taken from experiment.   Nevertheless, the framework is very significantly constrained and predictive.   From the underlying hypotheses of renormalizable local quantum field theory, and three generation structure, we derive that a $3\times 3$ unitary matrix, the CKM (Cabibbo, Kobayashi, Maskawa) matrix,  must describe a multitude of {\it a priori\/} independent decay rates and mixing phenomena, including several manifestations of CP violation.   

The first two Figures, one numerical and one graphic, give some sense of the rigor and power of these predictions.
 \begin{figure}
\begin{center}
\includegraphics[scale=.25]{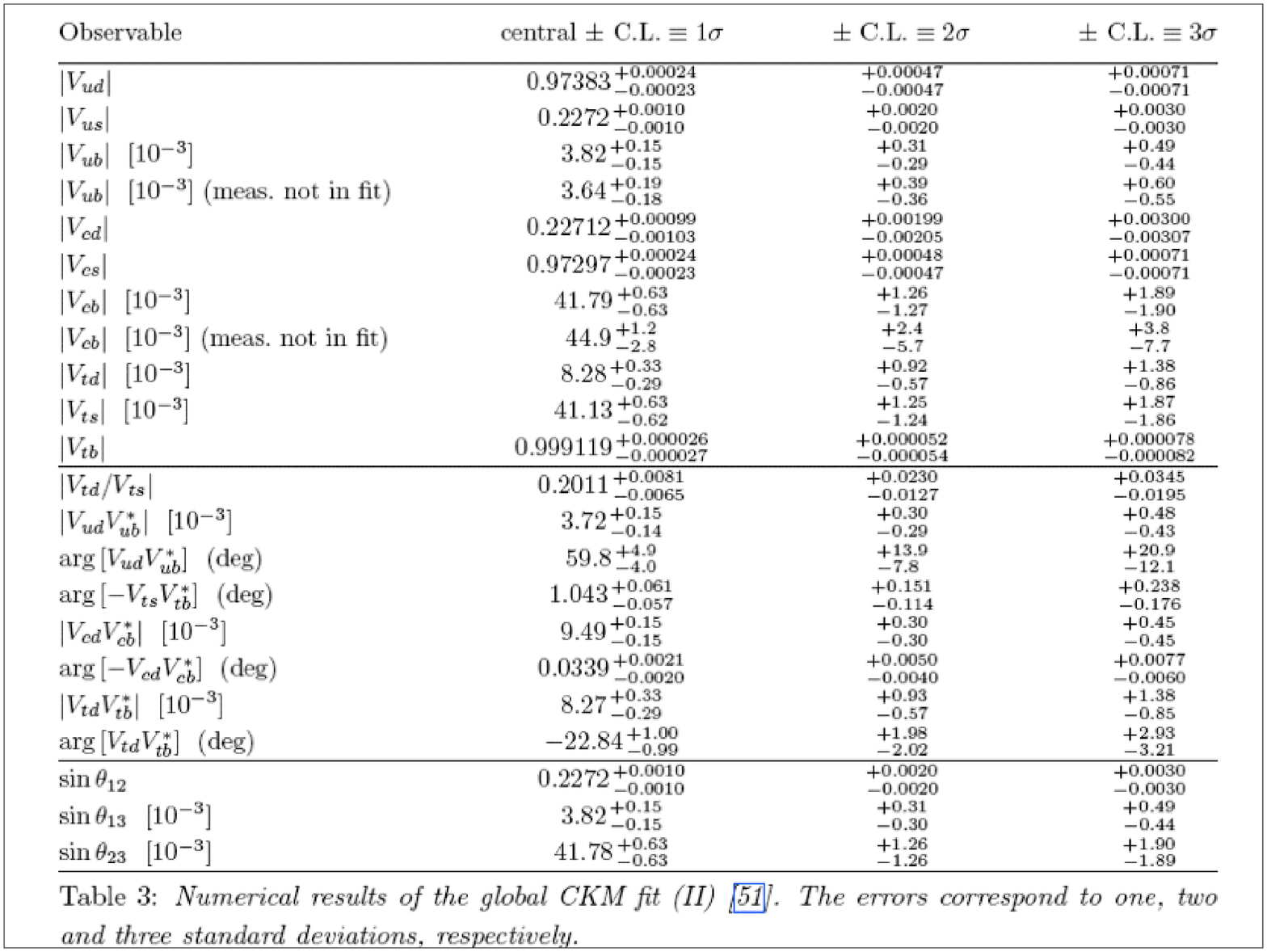}
\caption{A sampling of the quantitative results that test the ``ugly'' part of the standard model, that is its description of quark and lepton masses and mixings.   From \cite{flavorPaper}.}
\label{flavorData}
\end{center}
\end{figure}

\begin{figure}
\begin{center}
\includegraphics[scale=.40]{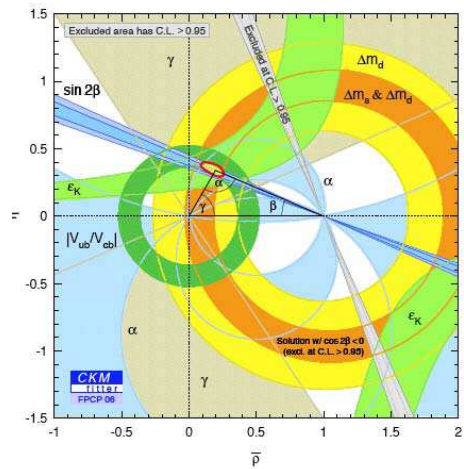}
\caption{Quantitative tests of the CKM framework, presented graphically.   From \cite{flavorPaper}.}
\label{flavorGraphic}
\end{center}
\end{figure}


Phenomena associated with neutrino masses, and with gravity, are commonly regarded as beyond, or at least outside, the standard model.    Of course, where one draws the boundary of the standard model is large\-ly a matter of taste.   But it's appropriate to emphasize that our working descriptions both of neutrino masses and of gravity fit smoothly and naturally into the conceptual framework associated with the ``core'' standard model of strong and electroweak interactions.    Specifically, neutrino masses can be accommodated using dimension 5 operators, and gravity through the Einstein-Hilbert curvature term and minimal coupling to matter (we can also include a cosmological term).    The deep guiding principles that underlie the standard model, to wit local quantum field theory based on operators of the lowest available mass dimension, also work to give theories of neutrino masses and of gravity that describe all existing observations in terms of a small number of parameters.   

Altogether, the standard model supplies an economical, precise and (we now know) extraordinarily accurate description of an enormous range of phenomena.   It supplies, in particular, firm and adequate foundations for chemistry (including biochemistry), materials science, and most of astrophysics.   We should be very proud of what we, as a community stretching across continents and generations, have accomplished.     

\subsection{An Unfinished Agenda}

But the success of the standard model, while imposing, is not complete.   The standard model has esthetic deficiencies, and there are phenomena that lie beyond its scope.    That combination of flaws is, ironically, full of promise.   We may hope that by addressing the esthetic deficiencies, we will bring in the phenomena.   I'll be discussing several  concrete examples of that kind.   

More generally, by drawing the boundaries of the known sharply, the standard model gives shape and definition to the unknown.   Here is an agenda of questions that arise out of our present combination of knowledge and ignorance:
\begin{itemize}
\item What drives electroweak symmetry breaking?
\item Do the gauge interactions unify? 
\item What about gravity? 
\item What is the dark matter? 
\item What is the dark energy?
\item How can we clean up the messy bits?
\item What else is out there?
\end{itemize}

\begin{figure}
\begin{center}
\includegraphics[scale=.50]{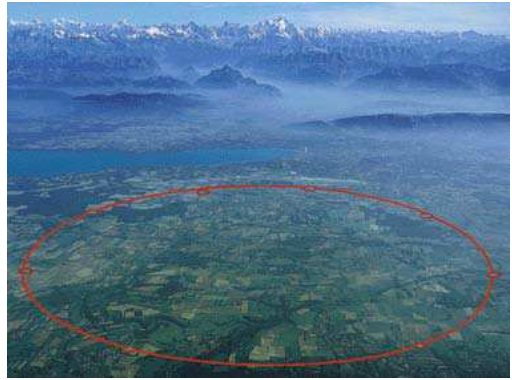}
\caption{The Large Hadron Collider: its scale and locale.  It is our civilization's answer to the ancient Pyramids, but better.  It is a monument to curiosity, not superstition; and to cooperation, not command.}
\label{LHCOver}
\end{center}
\end{figure}

\begin{figure}
\begin{center}
\includegraphics[scale=.25]{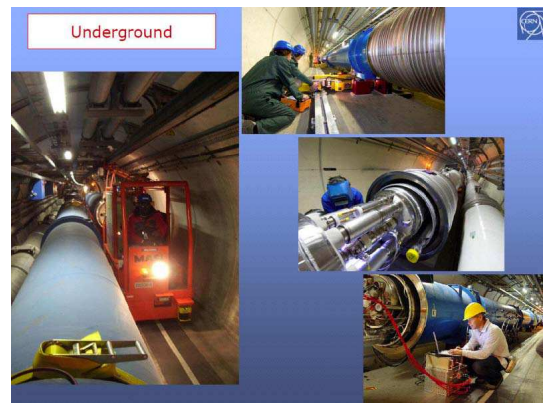}
\caption{The Large Hadron Collider: its interior.  Not only the scale, but especially the intelligence and precision of detail that go into this project lend it grandeur.}
\label{LHCUnder}
\end{center}
\end{figure}

Thanks to generous investment by the international community, and heroic work by many talented individuals, we will soon have a magnificent new  tool, the Large Hadron Collider (LHC), to address many of these questions.  (Figures \ref{LHCOver}, \ref{LHCUnder}.)

Of course, I don't know what we'll be finding, but I think it's possible to make some interesting guesses, and that is what I'll be doing here.  Returning to the agenda, I've highlighted in {\bf boldface\/} the questions that seem ripe for decisive progress, and in {\it italic\/} the questions I think are ripe for significant progress, and left the truly obscure questions plain.   These judgments emerge from the considerations that follow.

\begin{itemize}
\item {\bf What drives electroweak symmetry breaking?}
\item {\bf  Do the gauge interactions unify?} 
\item {\it What about gravity?} 
\item {\it What is the dark matter?} 
\item What is the dark energy?
\item How can we clean up the messy bits?
\item {\it What else is out there?}
\end{itemize}

\section{Electroweak Symmetry Breaking}

\subsection{The Cosmic Superconductor}

The success of the electroweak sector of the standard model teaches us that what we perceive as empty space is in reality a cosmic superconductor -- not, of course, for electromagnetic fields and currents, but for the currents that couple to $W$ and $Z$ bosons.   We do not know the mechanism or the substrate -- i.e., what plays the role, for this cosmic superconductivity,  that Cooper pairs play for ordinary metallic superconductivity.   No presently known form of matter can play that role, so there must be more.

\subsection{Minimal Model and Search}

The most economical assumption about what's missing, measured by degrees of freedom, is incorporated in the minimal standard model.   In this minimal model, besides the known fermion and gauge fields, we introduce a complex scalar $SU(2)$ doublet ``Higgs'' field.   Of the four quanta this complex doublet brings in, three have been observed: the longitudinal components of the $W^+, W^-$ and $Z$ bosons.  The remaining $\frac{1}{4}$ is the so-called Higgs particle. 

The Higgs particle has been a target of experimental search for many years now, and we'll be hearing much more about it in the next few days, so elaborate discussion is superfluous here.   Let me just present an icon (Figure \ref{higgsGraph}):

 \begin{figure}
\begin{center}
\includegraphics[scale=.35]{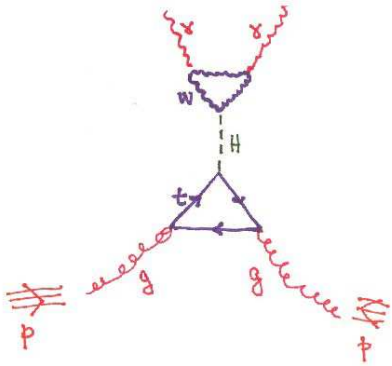}
\caption{A proposed mechanism for production and observation of the Higgs particle.  It is a purely quantum-mechanical process that brings in every portion of the standard model.}
\label{higgsGraph}
\end{center}
\end{figure}

This figure depicts an important search mode for the Higgs particle: production through gluon fusion, followed by decay into two photons.   The phrase ``Yesterday's sensation is today's calibration" conveys the pioneering ethos that is a glory of our community, but it is good on occasion to step back and appreciate how far we have come.   The description of this process, leading from colliding protons to resonant $\gamma \gamma$ production, brings in every sector of the standard model, including such profundities as the gluon structure of protons, the universal color coupling of QCD, the basic Higgs couplings proportional to mass, and the electroweak Yang-Mills vertex.   Moreover, this process is purely quantum-mechanical, twice over.   Each of its two loops indicates that a quantum fluctuation has occurred; interaction with virtual particles is essential both for the production and for the decay.    Yet we claim to understand this rare, involved, and subtle process well enough that we can calculate its rate, and distinguish it from many conceivable backgrounds.   It is an impressive calibration, indeed.  With any luck, it will become tomorrow's sensation!

\section{Unification and Supersymmetry}

\subsection{Unification of Charges}

The structure of the gauge sector of the standard model gives
powerful suggestions for its further development.  

\begin{figure}
\begin{center}
\includegraphics[scale=.42]{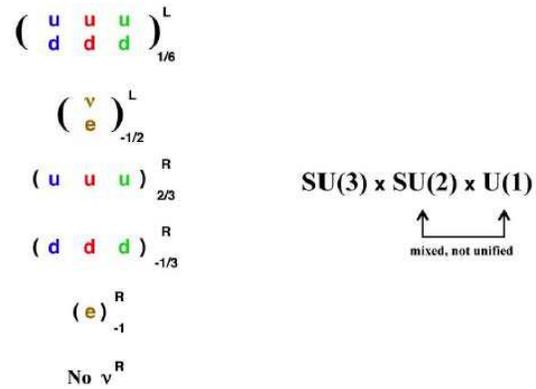}
\caption{The groups and multiplets of the standard model.   Esthetic defects: there are three separate gauge groups, and corresponding couplings; five independent fermion multiplets (not counting family triplication); and peculiar hypercharge assignments tailored to experiment.}
\label{smMultiplets}
\end{center}
\end{figure}

The salient features of the gauge sector of the standard model are displayed in Figure \ref{smMultiplets}.

The gauge theories of strong and electroweak interactions successfully describe a vast amount of data quantitatively, in terms of a very small number of input parameters (i.e., just three continuous ones).   Thus these theories are economical, as well as precise and accurate.    
They appear to represent Nature's last word, or close to it, on an enormous range of phenomena.

Yet there is room for improvement.   The product
structure $SU(3)\times SU(2) \times U(1)$, the reducibility of the fermion
representation, and the peculiar values of the hypercharge
assignments all suggest the possibility of a larger symmetry,
that would encompass the
three factors, unite the representations, and fix the hypercharges.
The devil is in the details, and it is not at all automatic that the observed,
complex pattern of matter will fit neatly into a simple mathematical structure.
But, to a remarkable extent, it does.  The smallest simple group into
which $SU(3)\times SU(2) \times U(1)$ could possibly fit, that is $SU(5)$,
fits all the fermions of a single family into two representations
($\bf{10} +\bar{\bf 5}$), and the hypercharges click into place. 

\begin{figure}
\begin{center}
\includegraphics[scale=.35]{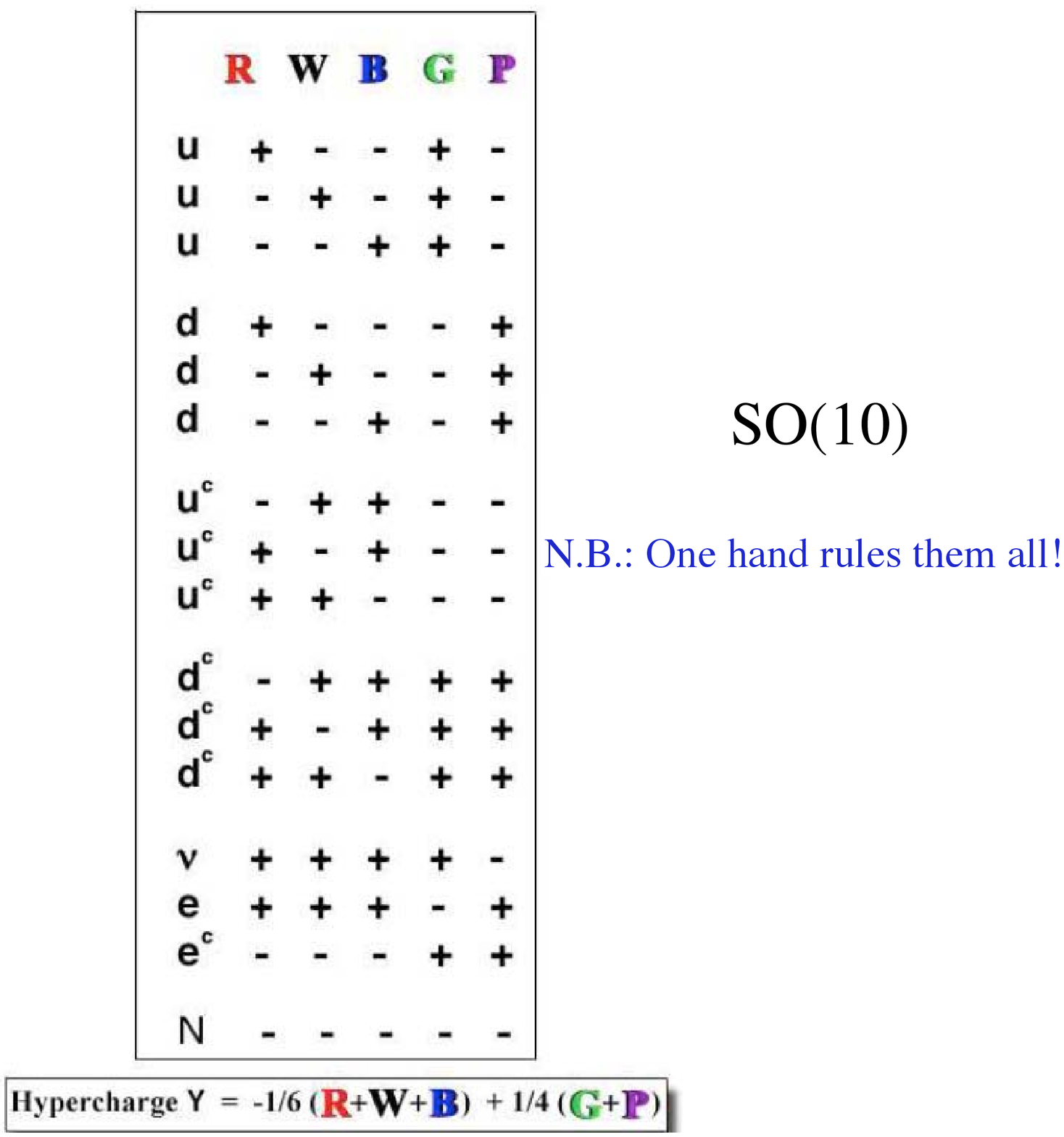}
\caption{The extended symmetry $SO(10)$ incorporates $SU(3)\times SU(2)
\times U(1)$ as a subgroup.   The five disparate fermion representations of the standard model gauge sector are united within a single internal-space spinor {\bf 16}.   The spinor contains one additional degree of freedom, which plays an important role in the theory of neutrino masses.  Hypercharge assignments are now related to weak and strong color charges. }
\label{so10Multiplets}
\end{center}
\end{figure}

As displayed in Figure \ref{so10Multiplets}, a
larger symmetry
group, $SO(10)$, fits these and one additional $SU(3)\times SU(2) \times
U(1)$ singlet particle into a single representation, the spinor $\bf{16}$.  
That additional particle is actually quite welcome.  It has the quantum
numbers of a right-handed neutrino, and it plays a crucial role in
the attractive ``seesaw" model of neutrino masses.  (See below, and for a more extended introduction to these topics see \cite{spacePart}.)

Perhaps most remarkably, according to this extended symmetry hypercharge assignments are no longer arbitrary and unrelated to the color and weak charges.   The formula
\begin{equation}
\label{hyperchargeFromColor}
Y ~=~ -\frac{1}{6} (R+W+B) + \frac{1}{4} (G+P)
\end{equation}
relating hypercharge $Y$, color charges $R,W,B$ and weak charges $G, P$ is a consequence of the symmetry.   With it, the loose end hypercharge assignments of the standard model join the central thread of unification.

\subsection{Unification of Couplings}

Unification of charges within $SO(10)$, or alternative (closely related) symmetry groups, displays a marvelous correspondence between the physically real and the mathematically ideal.   At first sight, however, its application to reality seems to fail quantitatively.    
For the unification of quantum numbers, though attractive, remains
purely formal until it is embedded in a physical model.  To do that, one must
realize the enhanced symmetry in a local gauge theory. But nonabelian gauge symmetry requires universality.  The $SO(10)$ symmetry requires that the relative strengths of the $SU(3)\times SU(2)\times U(1)$ couplings must be equal, which is not what's observed.

Fortunately, there is a compelling way to save the situation. If the higher symmetry is broken at a large energy scale (equivalently, a small distance scale),
then we observe interactions at smaller energies (larger
distances) whose intrinsic strength has been affected by the physics of
vacuum polarization, and those distorted couplings need not be equal.  
The running of couplings is an effect that can be
calculated rather precisely, in favorable cases (basically, for weak coupling).
Given a definite hypothesis about the particle spectrum, we get a definite prediction for the distortion of couplings, which we can compare with observation.  In this way
we can test, quantitatively, the idea that the observed couplings derive
from a single unified value.

\begin{figure}
\includegraphics[scale=.50]{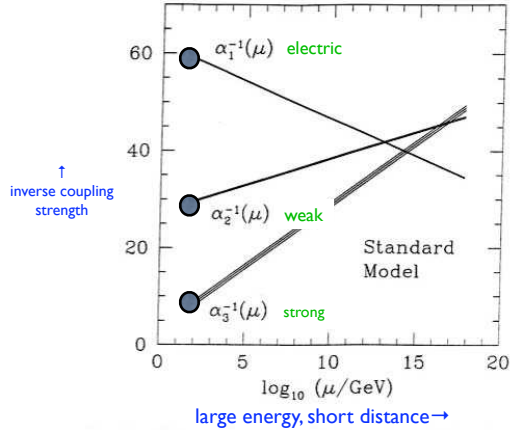}
\caption{Running of couplings, taking into account vacuum polarization due to standard model fields.  The width of the lines indicates the uncertainty in the measured values.}
\label{smRunning}
\end{figure}

Results from these calculations are tantalizing.
If we include vacuum polarization from the particles we know about in
the minimal standard model, we find approximate unification \cite{GQW}.  This is displayed in Figure \ref{smRunning}. 

\begin{figure}
\begin{center}
\includegraphics[scale=.38]{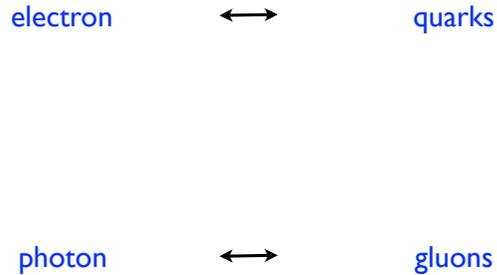}
\caption{Schematic indication of the unification of charge.   Gauge particles, notably including photons and gluons, are unified within a common representation, as are fermions, notably including electrons and quarks.}
\label{unificationOne}
\end{center}
\end{figure}

Though the general trends are encouraging, the different couplings do not become equal within the experimental uncertainty.
Were we to follow the philosophy of Sir Karl Popper, according to which the goal of science is to produce falsifiable theories, we could at this point declare victory.  For we turned the idea of unification of charge, schematically indicated in Figure \ref{unificationOne}, into a theory that was not merely falsifiable, but actually false.   

\subsection{Unification $\bf \heartsuit$ SUSY}

Our response is quite different: we are not satisfied with the hollow victory of falsification.  Having a beautiful idea that nearly succeeds, we look to improve it, by finding a still more beautiful version that works in detail.  We seek {\it truthification}.   

\begin{figure}
\begin{center}
\includegraphics[scale=.40]{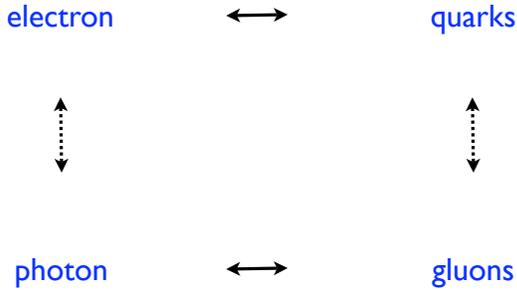}
\caption{Schematic indication of unification including SUSY.   Now particles of different spins fall into common multiplets.}
\label{unificationTwo}
\end{center}
\end{figure}

The central topic of this conference. of course, is  the possibility of another kind of symmetry, supersymmetry or SUSY, that enables further unification in the other direction, as indicated schematically in Figure \ref{unificationTwo}.

\begin{figure}
\begin{center}
\includegraphics[scale=.45]{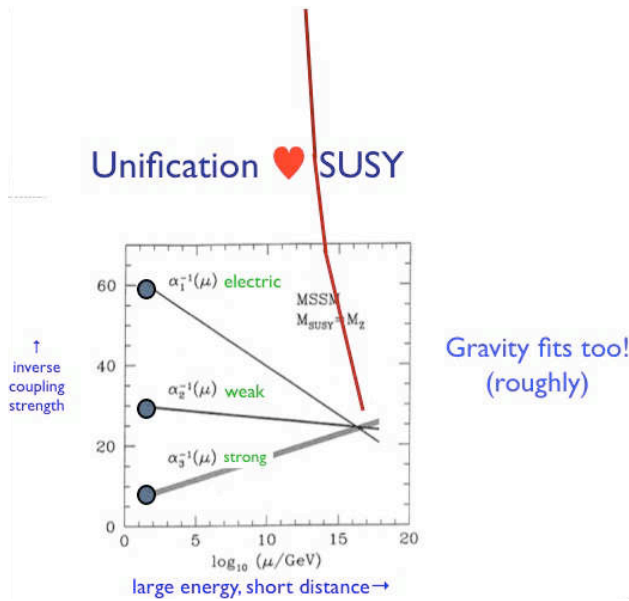}
\caption{Running of couplings, taking into account vacuum polarization due to all the fields involved in the minimal extension of the standard model to include supersymmetry, starting at a mass scale $\sim 1\  {\rm GeV}$.  The width of the lines indicates the uncertainty in the measured values.   Gravity runs classically.   It is ridiculously small at accessible scales, but we extrapolate to its (rough) equality with the other interactions at a common unification scale.}
\label{msmRunning}
\end{center}
\end{figure}

If we include
vacuum polarization from the particles needed to expand the standard
model to include supersymmetry, softly broken at the TeV scale, we find
accurate unification \cite{susyRunning}, as shown in Figure \ref{msmRunning}.   

Within this circle of ideas, called ``low-energy supersymmetry'', we predict the existence of a whole new world of particles with masses in the TeV range.  There must be supersymmetric partners of all the presently known particles, each having the same quantum numbers as known analogue but differing in spin by $\frac{1}{2}$, and of course with different mass.    Thus there are spin-$\frac{1}{2}$ gauginos, including gluino partners of QCD's color gluons and wino, zino, and photino partners of $W,Z,\gamma$, spin-0 squarks and sleptons, and more (Higgsinos, gravitinos, axinos).    Some of these particles ought to become accessible as the Large Hadron Collider (LHC) comes into operation.

On the other hand, many proposals for physics beyond the standard model
at the TeV scale (Technicolor models, large extra dimension scenarios, most
brane-world scenarios) corrupt the foundations of the unification of
couplings calculation, and would render its success accidental. 

\subsubsection{Importance of the Emergent Scale}

The unification occurs at a very large energy scale $M_{\rm unification}$, of
order $10^{16}$ GeV .  This success is robust against small changes in the
SUSY breaking scale, and is not adversely affected by incorporation of
additional particle multiplets, so long as they form complete representations
of $SU(5)$.

Running of the couplings allows us to infer, based entirely on low-energy data, an enormously large new mass scale, the scale at which unification occurs.    The disparity of scales
arises from the slow
(logarithmic) running of inverse couplings, which implies that
modest differences in observed couplings must be made up by a long interval of
running.  

The appearance of a very large mass scale is profound, and welcome on
several grounds:

\begin{itemize}
\item
Earlier we discussed the accommodation of neutrino masses and mixings within the standard model, through use of nonrenormalizable couplings.   With unification, we can realize those couplings as low-energy approximations to more basic couplings that have better high-energy behavior, analogous to the passage from the Fermi theory to modern electroweak theory.   

Indeed, right-handed neutrinos can have normal, dimension-four Yukawa couplings to the
lepton doublet. In $SO(10)$ such couplings are pretty much mandatory,
since they are related by symmetry to those responsible for
charge-$\frac{2}{3}$ quark masses.  In addition, since right-handed neutrinos are neutral under
$SU(3)\times SU(2)\times U(1)$
they, unlike the fermions of the standard
model, can have a Majorana type self-mass without violating those low-energy
symmetries.   We might expect the self-mass to arise where it is first
allowed, at the scale where $SO(10)$  breaks (or, in other models of unification, its moral equivalent).
Masses of that magnitude remove the right-handed neutrinos from the accessible
spectrum, but they have an important indirect effect. In second-order
perturbation theory the ordinary left-handed neutrinos, through their
ordinary Yukawa couplings, make virtual transitions to their right-handed
relatives and back.  (Alternatively, one substitutes 
\begin{equation}
\frac{1}{p \!\!\! / -M_{\nu_R}} ~\rightarrow ~ \frac{1}{-M_{\nu_R}}
\end{equation}
in the appropriate propagator.)
This generates non-zero masses for the ordinary
neutrinos that are much smaller than the masses of other leptons and quarks.

The masses predicted in this way are broadly consistent with
the tiny observed neutrino masses.  That is, the mass scale associated with the effective nonrenormalizable coupling, that we identified earlier, roughly coincides with the unification scale deduced from coupling constant unification.   Many, though certainly not all, concrete models of $SO(10)$ unification predict $M_{\nu_R} \sim M_{\rm unification}$.   No more than order-of-magnitude success
can be claimed, because relevant details of the models are poorly
determined.

\item
Unification tends to obliterate the distinction between quarks and leptons,
and hence to open up the possibility of proton decay. Heroic experiments
to observe this process have so far come up empty, with limits on partial
lifetimes approaching $10^{34}$ years for some channels. It is very difficult
to assure that these processes are sufficiently suppressed, unless the
unification scale is very large.  Even the high scale indicated by running of
couplings and neutrino masses is barely adequate. Interpreting this state of affairs
positively, we reckon that experiments to search for proton decay remain a most
important and promising probe into unification physics.
\item
Similarly, it is difficult to avoid the idea that unification brings in new
connections among the different families.  
Experimental constraints on strange\-ness-changing neutral currents and lepton
number violation are especially stringent.  These and other exotic processes that must be suppressed,
and that makes a high scale welcome.
\item
Axion physics requires a high scale of Peccei-Quinn (PQ) symmetry breaking,
in order to implement weakly coupled, ``invisible" axion models.  (See below.)  Existing observations only bound the PQ scale from below, roughly as $M_{\rm PQ} > 10^9\ {\rm GeV}$.   
Again, a high scale is welcome.  Indeed many, though certainly not all, concrete models of PQ symmetry suggest $M_{\rm PQ} \sim M_{\rm unification}$.    
\item
The unification of electroweak interactions with gra\-vity becomes much more plausible.
Newton's constant has units of mass$^{- 2}$, so it runs even
classically. Or, to put it less technically, because gravity responds directly to energy-momentum,
gravity appears stronger to shorter-wavelength, higher-energy probes. 

Because
gravity starts out extremely feeble compared to other interactions
on laboratory scales, it becomes roughly equipotent with them only at
enormously high scales, comparable to the Planck energy
$\sim 10^{18}\ {\rm GeV}$. This is not so different from $M_{\rm unification}$.   That numerical coincidence might be a fluke; but it's prettier to think that it betokens the descent of all these interactions from a common source.  Note that all these couplings have closely similar geometric interpretations, as measures of the resistance of fields (gauge or metric) to curvature.
\end{itemize}

\subsection{SUSY as Calibration}

If low-energy supersymmetry is a feature of our world, several of its particles will be discovered at the LHC.   That would, of course, be a great discovery in itself.    As we've seen, it would also be a most encouraging vindication of compelling suggestions for unification of the laws of physics, and for the bold extrapolation of the laws of quantum mechanics and relativity far beyond their empirical origins.  

But tomorrow's sensation is the day-after-tomorrow's calibration, and we can look forward to {\it using\/} the super-world as a tool for further exploration:
\vspace{-1ex}
\begin{itemize}
\item Some superpartner masses and couplings should, like the gauge couplings, derive from unified values distorted in calculable ways by vacuum polarization.  See Figure \ref{otherUnifications} and \cite{moreUnifications}.   Pursuing these relations could grow the ``one-off'' success of unification of gauge couplings into a thriving ecology.

\begin{figure}
\begin{center}
\includegraphics[scale=.43]{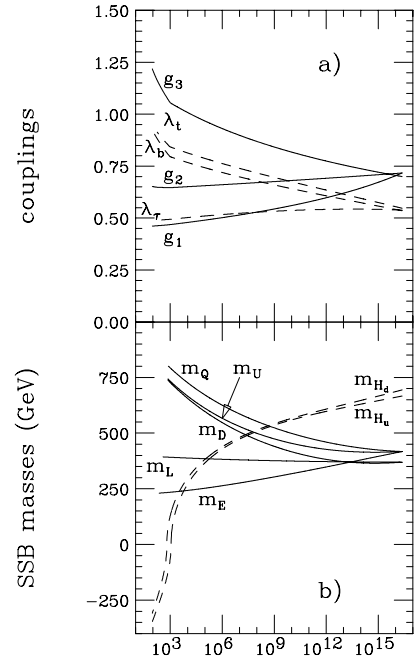}
\caption{Many models of low-energy supersymmetry incorporate sources of mass, or couplings, that are constrained by the unified gauge symmetry.   The logic of running gauge couplings likewise predicts numerical relations among the observed values of these masses and couplings \cite{baer}.}
\label{otherUnifications}
\end{center}
\end{figure}

\item There is currently no consensus regarding the mechanism of supersymmetry breaking.  It is a vast and unsettled subject, that I will not engage seriously here.   But as Figure \ref{susyBreaking} indicates, the leading ideas about the mechanism of SUSY breaking invoke exciting new physics, and predict distinctive signatures.   


\item Many model implementations of low-energy supersymmetry include a particle that is extremely long-lived, interacts very feebly with ordinary matter, and is abundantly produced as a relic of the big bang. Such a particle is a candidate to provide the dark matter that astronomers have observed.   This aspect deserves, and will now receive, a section of its own.  (For a wide-ranging review, see  \cite{khlopov}.)  

\begin{figure}
\begin{center}
\includegraphics[scale=.43]{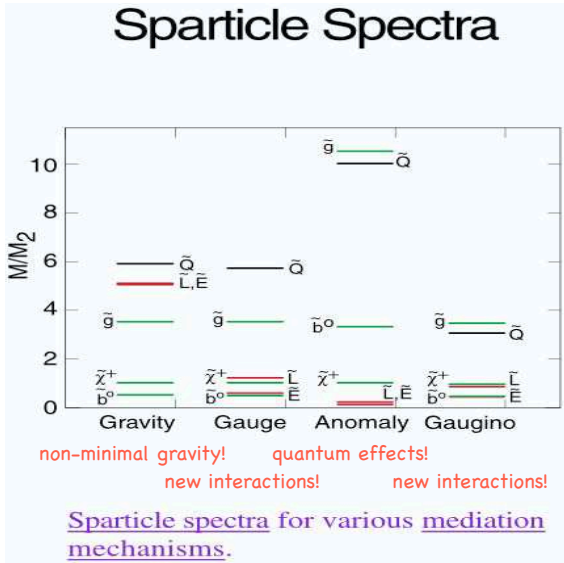}
\caption{Leading speculations about SUSY breaking invoke profound extensions of the known laws of physics.   They also predict distinctive spectra.   From \cite{susyBreaking}; see also \cite{moreSusyBreaking}.}
\label{susyBreaking}
\end{center}
\end{figure}

\end{itemize}

\section{Dark Matter}

\subsection{Dark Matter from Supersymmetry}

The multiplicative quantum number $ R \equiv (-1)^{3B+L+2S}$, where $B$, $L$, and $S$ are baryon number, lepton number, and spin respectively, is $+1$ for all standard model particles, and will be $-1$ for their superpartners.   Since $B, L$ and $S$ are to a very good approximation conserved, one expects that $R$ parity is to a very good approximation conserved.   Therefore the lightest $R$-odd particle is likely to be highly stable.   In many models of low-energy supersymmetry -- though, as we'll discuss shortly, by no means all -- this particle is stable on cosmological scales.  

\begin{figure}
\begin{center}
\includegraphics[scale=.43]{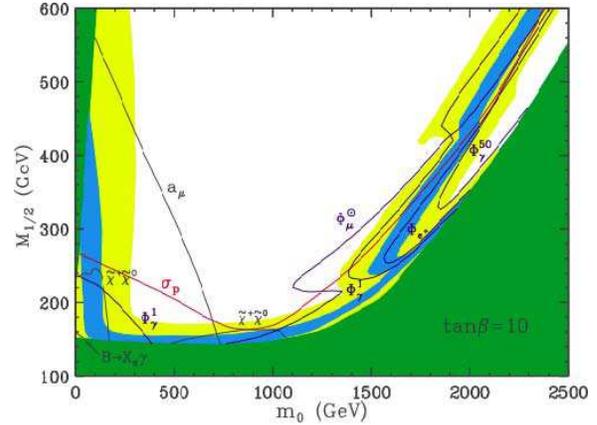}
\caption{Many model implementations of low-energy supersymmetry produce a dark matter candidate.   Here, from \cite{dmProduction}, is displayed the production of cosmologically stable particles in minimal supergravity-mediated models with $\tan \beta = 10$ and a range of universal scalar and gaugino masses $m_0, m_{\frac{1}{2}}$.   The green region is ruled out, either because the stable particle is electrically charged, or because it predicts charginos that would have been observed.   In the white region the dark matter is either overproduced (central region) or underproduced (sliver on the right).   In the yellow band a cosmologically interesting amount of neutralino dark matter is produced, in the blue band something close to the observed amount.   See \cite{dmProduction} for more details, and immediately below for some qualifications.}
\label{susyDark}
\end{center}
\end{figure}

Given a detailed model of low-energy supersymmetry, we can calculate the thermal history of the universe through the big bang, and estimate the surviving relic density of cosmologically stable particles.   Over a healthy range of parameters, the lightest $R$ odd particle is produced roughly in the right abundance to provide the dark matter, and also has the required property of interacting very feebly with ordinary matter.   This is illustrated in Figure \ref{susyDark}.   

\subsection{ ``Mission Accomplished''?}

It is all too easy to extrapolate a desired result from encouraging preliminary data.   

If low-energy supersymmetry and a dark matter candidate are discovered at the LHC, it will be a great enterprise to check whether the detailed properties of the particle, processed through the big bang, lead to the observed dark matter density.  That check is by no means a formality, because there are live, theoretically attractive alternatives to having the lightest $R$-parity odd particle observed at the LHC supply the cosmological density.   Both ``too much'' and ``too little'' SUSY dark matter are consistent with well-motivated, important physical ideas, as I will now explain.

\subsubsection{Too Much? - Superwimps}

There is a vast gap between decay lifetimes that can be detected at LHC and the age of the universe.   Thus a particle could {\it appear\/} to be stable at the LHC, and {\it calculated\/} to be abundant in the present universe, but absent in reality, because it decays on sub-cosmological time scales.  

That possibility is not as bizarre or contrived as it might sound at first hearing, for the following reason.   Superpartners of extremely feebly interacting particles, such as gravitinos or axinos, couple so feebly to ordinary matter that they are not accessible to observation at the LHC.   Yet they have odd $R$-parity.  Thus the lightest {\it observed\/} $R$-odd particle might well decay into a lighter {\it unobserved\/} $R$-odd particle with a lifetime that falls within the gap.   Since there is no good argument that the lightest standard model superpartner must be lighter than gravitinos, axinos, or other possible ``hidden sector'' particles (see below), this is a very much a live possibility. 

So if the dark matter candidate observed at the LHC appears to provide too much dark matter, or if it is electrically charged or otherwise cosmologically dangerous, a plausible interpretation is available.  

\subsubsection{Too Little? - Axions}

Given its extensive symmetry and the tight structure of relativistic quantum
field theory, the definition of QCD only requires, and only permits, a very
restricted set of parameters -- the quark masses, a coupling parameter, and one more, the $\theta$ parameter.  Physical results depend periodically upon
$\theta$, so that effectively it can take values between $\pm \pi$. The
discrete symmetries P and T are violated unless $\theta \equiv 0$ (mod $~\pi$).
We don't know the actual value of the
$\theta$ parameter, but only a limit, $|\theta | < 10^{-9}$.
Values outside this small range are excluded by experimental results,
principally the tight bound on the electric dipole moment of the neutron. 
Since there are P and T violating interactions in the world, the
$\theta$ parameter can't be set to zero by any strict symmetry.
So understanding its smallness is a challenge -- and an opportunity.

Peccei and Quinn discovered that if one
imposed a certain asymptotic symmetry, and if that symmetry is
broken spontaneously, then an effective value $\theta \approx 0$
results.
Weinberg and I explained that the approach
$\theta \rightarrow 0$ could be understood as a relaxation process,
whereby a very
light field, corresponding quite directly to $\theta$, settles into
its minimum energy state.
This is the axion field, and its quanta are called axions.

The phenomenology of axions is essentially controlled by one
parameter, $F$, with
dimensions of mass. It is the scale at which Peccei-Quinn symmetry breaks.

Axions, if they exist, have major cosmological implications, as I will now explain briefly.
Peccei-Quinn symmetry is unbroken at temperatures $T\gg F$.
When this symmetry breaks the initial value of the order parameter's phase is random beyond the then-current horizon scale. One can
analyze the fate of these fluctuations by solving the equations for a
scalar field in an expanding Universe.

The main general results are as follows.  There is an effective cosmic
viscosity, which
keeps the field frozen so long as the Hubble parameter
$H \equiv \dot{R} /R \gg  m$, where $R$ is the expansion factor and $m$ the axion mass.
In the opposite limit
$H \ll m$ the field undergoes lightly damped oscillations, which result in
an energy density that decays as $\rho \propto 1/R^3$.  Which is to say,
a comoving volume contains a fixed mass. The field can be regarded as a
gas of nonrelativistic particles in a coherent state, i.e. a Bose-Einstein condensate. There is some
additional
damping at intermediate stages. Roughly speaking we may say that
the axion field, or any scalar field in a classical regime, behaves
as an effective
cosmological term for $H>>m$ and as cold dark matter for $H\ll m$.
Inhomogeneous perturbations are frozen in while their length-scale exceeds
$1/H$, the scale of the apparent horizon, then get damped as they enter the horizon.

If we ignore the possibility of inflation, then there is a unique
result for the cosmic
axion density, given the microscopic model. The criterion $H \sim m$
is satisfied for
$T\sim \sqrt {\frac{M_{\rm Planck}}{F}}
\Lambda_{\rm QCD}$. At this point (and even more so at present) the horizon-volume contains many
horizon-volumes
from the Peccei-Quinn scale, but it still contains only a negligible
amount of energy
by contemporary cosmological standards. Thus in comparing to
current observations, it is appropriate to average over the starting
amplitude $a/F$ statistically.
If we don't fix the baryon-to-photon ratio, but
instead demand spatial flatness, as inflation suggests we should,
then $F \sim 10^{12}$ GeV correspond to the observed dark matter density, while for $F > 10^{12}$ GeV we get too much, and the relative
baryon density we infer is smaller than what we
observe.  

If inflation occurs before the Peccei-Quinn transition, this analysis
remains valid.
But if inflation occurs after the transition, things are quite different.

For if inflation occurs after the transition, then the
patches where $a$ is approximately homogeneous get
magnified to enormous size. Each one is far larger than the presently
observable Universe.
The observable Universe no longer contains a fair statistical sample of
$a/F$, but some particular ``accidental" value.  Of course there
is a larger region, which Martin Rees calls the Multiverse,
over which the value varies, but we sample only a small part of it.

Now if $F>10^{12}$ GeV, we could still be consistent with
cosmological constraints on the axion density, so long as
the starting amplitude satisfies
$ (a/F )^2 \sim ( 10^{12}~{\rm GeV})/F$.  The actual value of $a/F$,
which controls
a crucial regularity of the observable Universe, the dark matter density, is contingent in a very
strong sense. Indeed, it takes on other values at other locations in the multiverse.

Within this scenario, the anthropic principle is de\-monstrably correct
and appropriate \cite{linde}.
Regions having large values of $a/F$, in which axions by far dominate
baryons, seem likely to prove inhospitable for the development of complex
structures. Axions themselves are weakly interacting and essentially
dissipationless, and they dilute the baryons, so that these too stay dispersed.
In principle laboratory experiments could discover axions with $F >
10^{12}$ GeV.
If they did, we would have to conclude that the vast bulk of the
Multiverse is inhospitable to intelligent life. And we'd be forced to appeal
to the anthropic principle to understand the anomalously modest axion
density in our Universe.

\begin{figure}
\begin{center}
\includegraphics[scale=.300]{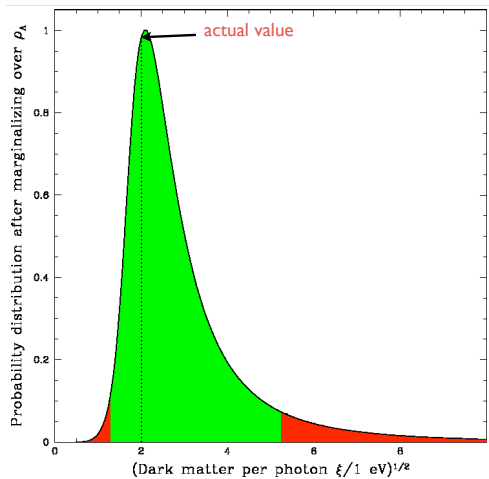}
\caption{In post-inflationary axion cosmology, the ratio of axion dark density to ordinary baryonic density varies over very large scales, and is subject to selection effects.   The prior is determined, as is the microphysics, so it is possible to take these effects into account with some semblance of rationality \cite{tegmark2}.   One finds that the dark matter density we actually observe is remarkably close to the most probable value.}
\label{axionDark}
\end{center}
\end{figure}

Though experiment does not make it compulsory, we are free to analyze the cosmological consequences of $F >> 10^{12}$ GeV.   Recently Tegmark, Aguirre, Rees and I carried out such an analysis \cite{tegmark2}.  We concluded that although the overwhelming {\it volume\/} of the Multiverse contains a much higher ratio of dark matter, in the form of axions, from what we observe, the typical {\it observer\/} is likely to see a ratio similar to what we observe.   See Figure \ref{axionDark}.

This post-inflationary axion cosmology is attractive in many ways.   It avoids the annoying axion string problem of traditional axion cosmology.  It relieves us of the necessity of bringing in a new scale: now $F$ could be the unification scale, or the Planck scale.   It has a potential cosmological signature, accessible in upcoming microwave background anisotropy measurements, as it provides a plausible source for isocurvature fluctuations that are larger in amplitude than gravitational wave fluctuations \cite{microwaveBackground}.   

And post-inflationary axion cosmology will happily generate the right amount of cosmological dark matter for us, if low-energy SUSY provides too little.  So if the dark matter candidate observed at the LHC appears to provide too little dark matter, and in particular if the lightest appreciably coupled R-odd particle decays into light species or into ordinary matter at a faster-than-cosmological rate, a plausible alternative source is available.

\section{Hidden Sectors and Portals}

\subsection{Might the LHC See Nothing?}

Let me begin the discussion of hidden sectors in what I trust will be a provocative way, with that question.  The usual answer is ``No, the LHC must discover new particles or new strong interactions in the vacuum channel, in order to avoid a crisis in quantum mechanics (loss of unitarity).''  The correct answer, however, is ``Yes,''   as I'll now demonstrate.      

\subsubsection{Division and Dilution}

Consider, to begin, adding to the standard model a singlet real scalar ``phantom'' field $\eta$.    All the couplings of gauge fields to fermions, and of both to the Higgs field, remain as they were in the original standard model.   This is enforced by gauge symmetry and renormalizability.   The Higgs potential is modified, however, to read
\begin{equation}
\label{phantomModifiedPotential}
V(\phi, \eta ) ~=~  - \mu_1^2 \phi^\dagger \phi + \lambda_1(\phi^\dagger \phi)^2 - \mu_2^2 \eta^2 + \lambda_2 \eta^4 - \kappa \phi^\dagger \phi \eta^2
\end{equation}
The only communication between $\eta$ and the standard model consistent with general principles is this $\kappa$ coupling to the Higgs field.   

The upshot of this simple cross-coupling is that when $\phi$ and $\eta$ acquire vacuum expectation values, the mass eigenstates (i.e., the observable particles) are created by mixtures of the conventional Higgs field and the phantom field.   The phantom component contributes nothing to the amplitude for production from conventional particles sources, i.e. quarks and gluons.   Thus the same overall production rate of Higgs particle is now divided between two lines.   Instead of finding a signal-to-noise ratio $S/N=2$, for the same exposure you'll get two channels with $S/N = 1$, which is not as good.

Of course, it's easy to generalize this model.   With more phantom fields, one has more division of strength.   And whereas 1 $5 \sigma$ signal is a discovery, as a practical matter 5 independent $1\ \sigma$ signals are worthless.   

It gets worse.   The phantoms might actually be the ``Higgs fields'' of an entire new sector, that has its own gauge fields and matter.   Then the Higgs-phantom mixtures might also decay into particles of the new sector, which are effectively invisible.   So not only is production divided, but also decay is diluted.   

These effects of division and dilution could easily render the Higgs sector effectively invisible, whilst barely affecting any other aspect of phenomenology.   

{\it The good news}: If we start from the minimal standard model, which contains just a single Higgs doublet field -- and thus, after electroweak symmetry breaking, just a single neutral scalar -- the Higgs portal might be quite challenging to exploit.   Given a richer Higgs sector, including charged fields (as in low-energy SUSY), or more if ``Higgs particles'' appear as decay products of particles that are more identifiable, it could be much easier.   In any case, a new world would be open to exploration.

\subsubsection{An Example: Mass from Quantum Mechanics}

Let us add, in the spirit of counting, an $SU(4)$ symmetry to the standard model, so that the gauge group becomes 
$$
G ~=~ SU(4)\times SU(3)\times SU(2)\times U(1)
$$
In the spirit of coupling unification, we suppose that $SU(4)$ is a super-strong interaction.   It can support spontaneous chiral symmetry breaking, which we represent by some sort of $\sigma$ model.  In the simplest version the new $\vec \sigma$ field is a 4-component vector; but it could also be some more elaborate matrix.  Assuming all the fields charged under $SU(4)$ are $SU(3)\times SU(2)\times U(1)$ singlets, the important modification to the standard model will come through a modified effective potential, generalizing Eqn. (\ref{phantomModifiedPotential}):
\begin{equation}
\label{scaleModelModifiedPotential}
V(\phi, \eta ) ~=~  - \mu_1^2 \phi^\dagger \phi + \lambda_1(\phi^\dagger \phi)^2 - \mu_2^2 {\vec \sigma}^2 + \lambda_2 ({\vec \sigma}^2)^2 - \kappa \phi^\dagger \phi {\vec \sigma}^2
\end{equation}

The $\kappa$ coupling will induce mixing, as before.   The non-Goldstone field $\sigma_0$, that encodes the magnitude of $\vec \sigma$, will decay into the massless phantom Nambu-Goldstone ``pions''.   So we get dilution, as well.   

It is entertaining to imagine $\mu_1^2 = 0$.   Then we have an underlying model in which there is no classical mass parameter anywhere.   Electroweak symmetry breaking is induced from nonperturbative, intrinsically quantum-mechanical chiral symmetry breaking in the phantom sector, through the cross-coupling $\kappa$.   In this indirect way we implement the vision that inspires technicolor models, while avoiding the usual phenomenological difficulties of such models.  Those difficulties arise because the new strongly interacting sector is not hidden (that is, if it does not consist of $SU(3)\times SU(2) \times U(1)$ singlets), and more specifically because the super-strong condensate itself breaks $SU(2)\times U(1)$.

\subsection{Motivations for Hidden Sectors}

The two little models we've discussed are not uninteresting in themselves.  Moreover, they illustrate possibilities that are more broadly motivated.   Here are some other reasons to consider the possibility of hidden sectors seriously:
\begin{description}
\item[Hippocratic oath: ]  At the opening of their Hippocratic Oath, prospective doctors  promise to ``abstain from whatever is deleterious and mischievous".   Hidden $SU(3)\times SU(2)\times U(1)$ singlet sectors, unlike many other speculative extensions of the standard model, do little harm.   They do not spoil the successful unification of couplings, nor do they open a Pandora's box of flavor violation.
\item[stacks and throats:]  In string theory, hidden sectors easily arise from far-away (in the extra dimensions) stacks of D-branes or orbifold points.  The original $E_8\times E_8$ heterotic string contains an early incarnation of a hidden sector.
\item[plays well with SUSY:]  Hidden sectors are invoked in several mechanisms of SUSY breaking.   And the next-to-minimal supersymmetric standard model (NMSSM), which introduces an extra  $SU(3)\times SU(2)\\ \times U(1)$ singlet chiral superfield, has been advocated on phenomenological grounds.  It eases some ``naturality'' problems.  
\item[flavor and axions:]  It is tempting to think that the complicated pattern of quark and lepton masses and mixings reflects a complicated solution to simpler basic equations; specifically, that the more fundamental equations have a flavor symmetry, which is spontaneously broken.   Phenomenology seems to require that flavor-symmetry breaking dynamics occurs at a high mass scale.  Therefore the order parameter fields must be $SU(3)\times SU(2)\times U(1)$ singlets, and they constitute a hidden sector in our sense.   Axion physics embodies this idea in a compelling way for one aspect of the quark mass matrix, i.e. its overall phase.     
\end{description}

\subsection{Bringing Method to the Madness}

Possible forms of communication between hidden sectors can be considered more abstractly, in the style of effective field theory.   We seek low-dimension operators suitable for inclusion in the world-Lagrangian that contain both standard model and hidden sector fields.   We assume these operators must be gauge and Lorentz invariant.    The simplest cases correspond to coupling in spin 0, $\frac{1}{2}$, or 1 hidden sector fields, building up dimension 4:
\begin{description}
\item[Spin 1:] An $SU(3)\times SU(2) \times U(1)$ vector $V^\mu$ can couple in three different ways to make a dimension 4 invariant operators.  It can couple to fermion $\bar f \gamma_\mu f$ currents.   This possibility has been much discussed under the rubric ``$Z^\prime$ bosons''.   It can couple to the hypercharge gauge curvature $B_{\mu \nu}$ through the current $\overleftarrow{ \partial}^\nu B_{\mu \nu}$.   This gives ``kinetic mixing''.   (It might also couple through the dual current $\overleftarrow{ \partial}^\nu \tilde{B}_{\mu \nu}$, to give a form of $\theta$-parameter mixing, but this appears to be of little consequence.)   Finally, $V^\mu$ might couple in through the Higgs field, appearing within the covariant derivative in $(\nabla^\nu \phi^\dagger)(\nabla_\nu \phi)$.   As the Higgs field condenses, this leads to mixing between $V^\mu$ and standard model gauge bosons at the level of mass eigenstates.   In general, in both kinetic and mass mixing, the hidden sector particles will acquire electric charges that need not be commensurate with the familiar unit (and presumably must be much smaller) .   
\item[Spin $\frac{1}{2}$:] An $SU(3)\times SU(2) \times U(1)$ spin-$\frac{1}{2}$ fermion $\xi$ can make a dimension 4 invariant operator by coupling in to the dimension $\frac{5}{2}$ singlet $\phi^\dagger L$, where $L$ is a left-handed lepton doublet.   If such an interaction occurs with a very small coefficient, it leads to a massive Dirac neutrino; if the coefficient is moderate, but $\xi$ has a large intrinsic mass, we integrate out $\xi$ to get the familiar see-saw mechanism to generate small physical neutrino masses. 
\item[Spin 0:]  An $SU(3)\times SU(2) \times U(1)$ spin-$\frac{1}{2}$ spin 0 particle can couple in through the dimension 2 singlet $\phi^\dagger \phi$.   This opens the Higgs portal to the hidden sector, as we discussed above.
\end{description}

Evidently this framework helps to organize several old ideas, and puts the Higgs portal idea in proper context.   

\section{Summary and Conclusions}

With the LHC, we will expand the frontiers of fundamental physics.

\begin{itemize}
\item We will learn, through a {\it tour de force\/} of  physics, what makes empty space function as a cosmic superconductor. 
\item We will learn whether existing indications for unification and supersymmetry have been Nature teaching us or Nature teasing us.
\item If indeed the superworld opens up, it will probably supply a good candidate for the dark matter.
It will then be a great enterprise to establish or disprove that candidate.  
\item Hidden sectors are entirely possible.  They could complicate things in the short run, but would teach us even more in the long run.  
\end{itemize}

It will lead to a new Golden Age, that could also be enriched by discoveries in precision low-energy physics (elementary electric dipole moments), rare processes (proton decay), and cosmology (primordial isocurvature or gravitational wave fluctuations).   

Given the available time and bandwidth, I've had to be very selective in my choice of topics, and sketchy in my discussion even of the topics chosen.  Still, I hope I've given you a sense of some of the ambitious issues and ideas that we can expect to advance dramatically in the next few years.%

\doefunds


\end{document}